# EXPLORING THE VIABILITY OF FISHER DISCRIMINANTS IN GALAXY MORPHOLOGY CLASSIFICATION

Sazatul Nadhilah Zakaria[1], Santtosh Muniyandy[1] & John Y. H. Soo[1*]

[1]*School of Physics, Universiti Sains Malaysia, 11800 USM, Pulau Pinang, Malaysia*

*\*Corresponding author: johnsooyh@usm.my*

**Abstract**

One of the major challenges in astronomy involves accurately classifying galaxies, particularly distinguishing between different galaxy types. While many complex algorithms have shown strong performance in classification tasks, their complexity often results in longer processing times and increased difficulty in understanding. This study addresses this issue by exploring the viability of Fisher discriminants, a much simpler algorithm, in performing galaxy morphology classification. We tested four machine learning algorithms: the Fisher discriminant, Artificial Neural Networks (ANNs), Boosted Decision Trees (BDTs), and $k$-Nearest Neighbours ($k$NNs) to classify galaxies by the shape of their central bulges. Using data from the Sloan Digital Sky Survey (SDSS), we utilised five pre-processing transformations: normalisation, decorrelation, principal component analysis (PCA), uniformisation, and Gaussianisation, and classified the shape of central bulge into either rounded or no-bulge, based on the Galaxy Zoo Decision Tree. When compared to the Galaxy Zoo 2 (GZ2) labels, the Fisher discriminant with uniformisation obtained the highest accuracy score of 0.9310, outperforming ANN, BDT, and $k$NN by 1.93%, 0.42%, and 3.08%, respectively.

## Introduction

Galaxies are vast systems made up of intergalactic gases, dust, stars, and planets, where these elements are bounded together by gravity. Galaxies are usually classified according to the Hubble Sequence (Hubble 1926), in the shape of a tuning fork, where they are divided into spirals and ellipticals. A spiral galaxy is made up of four components: (1) a giant rotating pinwheel with a system of spiral arms that can be wounded tightly or loosely and filled with younger stars, (2) a pancake-like disc of older stars, (3) a central bulge of tight concentrated stars, and (4) a surrounding halo of older stars. As for an elliptical galaxy, it does not have structured internal features, and the stars orbit its core in random directions. Having only a little gas and dust, stars in elliptical galaxies tend to be older as they do not have enough gas for star formation. It is theorised that this galaxy type emerges from the collisions and mergers of spirals, making ellipticals less common than the spiral galaxies (Burkert et al. 2003). Other classifications include the lenticular galaxies, which are combinations between spirals and ellipticals, having a central bulge and disc but no spiral arms; while irregular galaxies are those with odd shapes and do not have structured features. Classifying galaxies based on their properties is essential to gain insights into their dynamics and to map the universe's structure (Buta 2011).

Many recent sky surveys have been deployed to collect galaxy data across vast regions of space using advanced telescopes, they gather various morphological information of galaxies in different wavelengths of light and analyse them using specialised algorithms and tools. For example, Stoppa et al. (2023) developed the AutoSourceID-Classifier (ASID-C), a tool designed to provide a robust and effective solution for star-galaxy separation using the images captured by the MeerKAT's Optical Eye (MeerLICHT, Paterson 2017), an optical telescope located in Sutherland, paired with morphological classifications from the Dark Energy Camera Legacy Survey (DECaLS, Dey et al. 2019), with a particular focus on those with signal-to-noise (S/N) ratio near the detection limit. They found that ASID-C consistently outperforms star-galaxy classification results generated by both SourceExtractor (SE, Bertin et al. 1996) and a baseline model across most S/N values and metrics, particularly in the low S/N ratio regions where classifying stars and galaxies become more challenging. This suggests that their classifier generates more reliable and well-calibrated probability predictions. On the other hand, von Marttens et al. (2023) used supervised machine learning (ML) to automate a Tree-based Pipeline Optimization Tool (TPOT, Le et al. 2020) to find an optimised pipeline to perform star-galaxy-quasar classification, providing a value-added catalogue (VAC) with the best classifications for 47.4 million sources. They found that eXtreme Gradient Boosting (XGBoost, Chen et al. 2016) is the most suitable pipeline, achieving an average precision of $> 0.99$ for galaxies and stars, and $> 0.96$ for quasars, outperforming SE and the Stellar-Galaxy Loci Classifier (SGLC, López-Sanjuan et al. 2019).

One of the most notable data sets used to verify galaxy morphology classification is the Galaxy Zoo (Lintott et al. 2008), a citizen-science project where volunteers are invited to classify galaxies based on their shapes and features. Galaxy Zoo 2 (GZ2, Willett et al. 2013) is the second incarnation of Galaxy Zoo, it provided detailed morphological classifications for 304 122 largest and brightest galaxies from SDSS. Urechiatu et al. (2024) categorised 10 000 galaxy images obtained from GZ2 into five classes: completely round and smooth, in-between smooth, cigar-shaped smooth, edge-on, and spirals, using a newly designed convolution neural network (CNN)-based model. They performed data enhancement to prevent overfitting and compared their results with several neural networks like DenseNet (Huang et al. 2017), EfficientNet (Tan et al. 2019), MobileNet (Howard et al. 2017), and the method of Zhu et al. (2019). They found that their proposed model achieved an improvement in terms of accuracy and precision as high as 1.7% as compared to other models. Currently their model has the most reliable confidence estimates for galaxy morphology classification for GZ2, and it solely relied on the pixel information that the CNN can extract without performing additional operations. GZ2 data has also been used in several other galaxy morphology classification work like Beck et al. (2018), Reza (2021) and Vavilova et al. (2021), each using different classification methods, machine learning algorithms and sampling strategies. Many among them emphasised that the classification performance depends on the size and representation of the data set.

Throughout the years, machine learning has demonstrated significant success in solving astronomical problems, notably reducing the need for human intervention and increasing efficiency. The versatility of machine learning offers countless effective solutions, reflecting its inherent nature. From the research papers mentioned, it is observed that most studies utilise complex algorithms, often tested on large sample sizes ranging from hundreds of thousands to even millions of data points. These methods have been proven to deliver strong performance, but they come with certain drawbacks such as longer training time, difficult to understand, and taking up more storage. In contrast, simpler machine learning methods, such as the Fisher discriminant (Fisher 1936) is faster to run, easier to understand, and require less storage, making it a compelling alternative. The Fisher discriminant is a method developed by Ronald Fisher which identifies a linear combination of features to maximise class separation, by minimising the within-class variance while maximising the between-class variance. Raichoor et al. (2015) found it effective in selecting emission-line galaxies for the eBOSS survey which highlights its ability to target the specific galaxy populations with high accuracy. However, Fraix-Burnet et al. (2023) noted its limitations when dealing with complex datasets in galaxy morphology classification, suggesting that advanced discriminants such as neural networks often outperform it. These studies suggest that while Fisher discriminant can be successful for simple problems, they may struggle with more complex data.

Testing the performance of the Fisher discriminant in classifying galaxies could provide valuable insights into its efficiency and potential in comparison to the many current and complex approaches. Therefore, in this project, we aim to achieve the following objectives:

1. To test the viability of Fisher discriminant in galaxy morphology classification by comparing it with the performances of ANNs, BDTs and $k$NNs; and
2. To study the impact of pre-processing variable transformations on each of the classifiers used and evaluate their accuracies.

This paper is structured as follows. In the next section, the data and methodology performed in this study will be discussed. The concepts behind the software used (ANNz2), the data sets (SDSS, GZ2), as well as the magnitude fits and morphological parameters used in this study will be explained. Then, the four machine learning algorithms used: Fisher discriminants, ANNs, BDTs, and $k$NNs will be discussed, followed by an introduction to the five pre-processing transformations used in this study and the metrics used to evaluate the algorithm performance. The section following presents the results obtained in this study, and finally, the last section concludes this paper, and the possible future work and outlook will also be mentioned.

## Materials and Methods

### Data Sets

This paper uses data from the Sloan Digital Sky Survey (SDSS) and Galaxy Zoo 2 (GZ2), focusing on the branch describing the shape of the central bulge in the GZ2 decision tree, summarised in **Figure 1**. SDSS is one of the most influential astronomical surveys ever conducted, where it began its operation in 1998 with its first phase named SDSS-I (York et al. 2000). Recently, it has released DR18, which is part of its fifth phase, SDSS-V (Almeida et al. 2023). SDSS uses a *ugriz* photometric system (Doi et al. 2010; Fukugita et al. 1996), that consists of these five broad bands that capture light in specific wavelength. The use of such very wide bandpasses ensures high efficiency for faint object detection and allows coverage of the entire accessible optical wavelength range. The filters are integrated into charge-coupled devices (CCDs) within the SDSS photometric camera (Gunn et al. 1998), a component of the 2.5 m SDSS telescope (Gunn et al. 2006). SDSS provides photometric, morphologic and spectroscopic data, representing detailed optical images and properties of the objects.

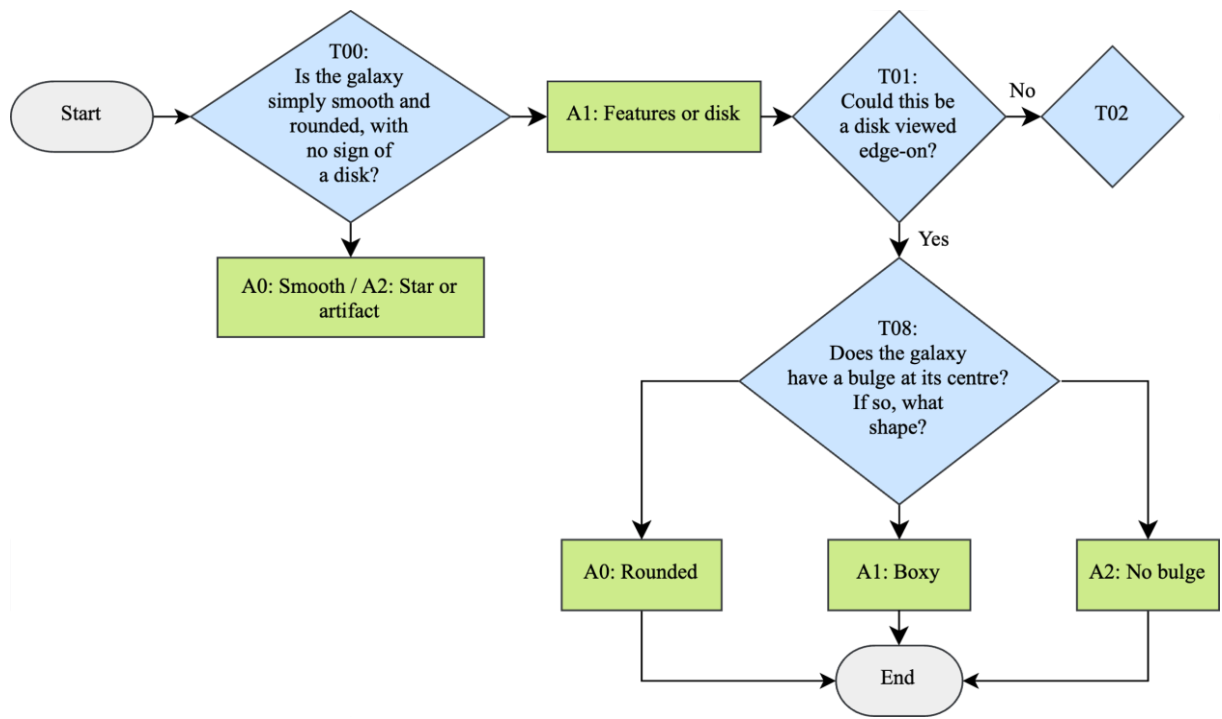

**Figure 1**. The methodological flowchart summarising a branch within the Galaxy Zoo 2 Decision Tree used in this work. The full decision tree can be found at https://data.galaxyzoo.org.

Meanwhile, GZ2 (Willett et al. 2013) is one of the many citizen science projects performed by Zooniverse[1], which is an online platform that allows volunteers from all over the world to participate in scientific research in various fields. Zooniverse utilises human's intuition and personal experiences to provide depth and complexity in decision-making processes, which often surpass machine learning capabilities. This approach helps to accelerate scientific discoveries and makes research accessible to a wider audience. In short, users using the GZ2 portal are shown randomly selected SDSS images of galaxies and asked to classify them by choosing from a set of possible answers, following 11 classification tasks outlined in the Galaxy Zoo Decision Tree[2]. This project completed in 14 months and produced a data set that comprises of 16 340 298 galaxy classifications made by 83 943 volunteers (Willett et al. 2013). Following GZ2, the Galaxy Zoo project expanded to incorporate diverse data sets from both observational and simulated sources, producing other surveys including Galaxy Zoo 3 Hubble (GZ3 Hubble, Willett et al. 2016), Galaxy Zoo Cosmic and Near-infrared Deep Extragalactic Legacy Survey (GZ4 CANDELS, Simmons et al. 2016), and Galaxy Zoo 4 Illustris (GZ4 Illustris, Dickinson et al. 2018), all of which have been completed.

---

[1] https://www.zooniverse.org.

In this work, we obtained the GZ2 morphology classification data (particularly the data which classifies a galaxy's central bulge into 'rounded' or 'no-bulge') and cross-matched this sample via object ID with the photometric data from SDSS to produce a data set of 1530 galaxies. The morphological data are filtered to produce a clean sample with `flag=1`.

*Input Variables and Training Size*

In every machine learning problem, it is crucial to determine the input variables (which are used to train the machine) and the output variable (which is the parameter we want to predict). In this work, we used 11 SDSS photometric parameters as inputs, all obtained using the *r*-band due to its low uncertainty values as compared to the others. The 11 variables are

- Four *apparent magnitudes*, measured using the point-spread function for stars (PSF, `psfMag_r`), fixed spectroscopic fibre size (`fiberMag_r`), galaxy model fit (`modelMag_r`) and Petrosian fit (`petroMag_r`);
- Three *Petrosian apparent radii*, encompassing the effective radius (`petroRad_r`), the radius at 50% light (`petroR50_r`), and the radius at 90% light (`petroR90_r`);
- Three *model fit probabilities*, describing the log likelihood of the object being a star (`lnLStar_r`), a spiral galaxy fitting an exponential model (`lnLExp_r`), and an elliptical galaxy fitting a de Vaucouleurs model (`lnLDeV_r`); and
- An *adaptive shape measure* (`mE1_r`), which describes the ellipticity of the object.

For more information on the morphological parameters mentioned above, the reader could refer to Stoughton et al. (2002) for full details. We have specially selected these 11 parameters as inputs to classify the galaxy's central bulge shape as we believe that the different model fits on apparent magnitude, radius and ellipticity may play an important role in determining the outcome. For example, a galaxy with a significantly rounded bulge may fit the galaxy exponential model better, thus showing a large difference between `lnLDeV_r` and `lnLexp_r`, as well as difference in values in the percentage light Petrosian radii `PetroR50_r` and `PetroR90_r`. The output parameter of this work is a flag where if its value is closer to 1, it is a galaxy with a rounded bulge, while closer to 0 implies no bulge is found. We used the data classified by the public in GZ2 as our data set, where we filtered the data to include only galaxies with clear and well-defined properties, i.e. `t09_bulge_shape_a25_rounded_flag=1` for rounded bulge, and `t09_bulge_shape_a27_no_bulge_flag=1` for no-bulge.

---

[2] Available at https://data.galaxyzoo.org under the section *Data Visualizations*.

Further filtering the data and removing bad input and output values, we obtained a sample of 1170 rounded galaxies and 360 no-bulge galaxies from SDSS DR8, giving a total of 1530 galaxies. We note that originally according to the GZ2 Decision Tree, there exists another category called 'boxy' galaxies which were classified in tandem with the rounded and no-bulge galaxies, however, the classification of boxy galaxies was omitted from this work due to insufficient data (there were only 6 galaxies available to be trained). This is probably because the observed galaxies from that branch are predominantly elliptical or lenticular types, which are known to be observed less frequently.

Due to the imbalanced number of galaxy types, we decided to create a data set of equal number of rounded and no-bulge galaxies, at 360 galaxies each, which gives a total of 720 galaxies. This data set is divided into the ratio of 2:2:1 for training, validation, and evaluation processes. The validation set is used to prevent overtraining, while the evaluation set is used to evaluate their performance. The signals and backgrounds were determined using a threshold of 0.5, meaning if the output flag results with a probability of $\geq 0.5$ are labelled as a signal (i.e. a galaxy with a rounded bulge), while labelled as a background if otherwise.

*ANNz2: the Fisher Discriminant and Other Machine Learning Algorithms*

ANNz2[3] is a machine learning software package that provides estimated solutions and probability distribution functions (PDFs) for regression and classification problems. Coded in python by Sadeh et al. (2016), ANNz2 uses the Toolkit for Multivariate Data Analysis (TMVA) package[4] which was integrated into the ROOT C++ software framework (Brun et al. 1997). ANNz2 can be run in five configurations: single regression, randomised regression, and binned classification for regression problems, while single classification and randomised classification for general classification problems. ANNz2 version 2.3.2 was setup on a Linux machine running on Ubuntu 23.10, with ROOT 6.30.02 and python3 installed.

We have selected to use ANNz2 as it allows us to run the Fisher discriminant method on our data, which is the machine learning algorithm of our interest to be experimented on. To assess its viability, we compared its performance with three other machine learning algorithms, namely artificial neural networks (ANNs), boosted decision trees (BDTs) and $k$-nearest neighbours ($k$NNs), to be used as benchmarks to evaluate the performance of the Fisher discriminant. These four machine learning algorithms are described in the following paragraphs.

---

[3] The code is available for download at https://github.com/IftachSadeh/ANNZ.
[4] Further information at https://root.cern/manual/tmva/.

A discriminant analysis is essentially a method which finds the linear combination of input parameters to separate objects in two or more classes. In the case of the Fisher discriminant (Fisher 1936), we let $x_k(i)$ be the $k$-th input variable of object $i$, and $y_F(i)$ be the output classification of object $i$. Assuming that in our training sample we have $N_S$ signals and $N_B$ background objects (i.e. in our case, $N_S$ rounded and $N_B$ no-bulge objects), the Fisher discriminant method will yield

$$y_F(i) = F_0 + \sum_{k=1}^{n_{\text{var}}} F_k x_k(i), \qquad (1)$$

where $F_k$ is the Fisher coefficient, one for each input variable; and $F_0$ a constant which centres the sample mean $\bar{y}_F$ of all training objects $N_S + N_B$ at zero. The Fisher coefficients for each variable $k$ can be written as

$$F_k = \frac{\sqrt{N_S N_B}}{N_S + N_B} \sum_{l=1}^{n} W_{kl}^{-1} (\bar{x}_{S,l} - \bar{x}_{B,l}), \qquad (2)$$

which is in the form of the sum of a multiplication between the difference between the sample means of the signal and the background, and the inverse of a within-class matrix $W_{kl}$, which is given by

$$W_{kl} = C_{S,kl} + C_{B,kl} = \sum_{k,l}^{N_S} (x_{S,k} - \bar{x}_{S,k})(x_{S,l} - \bar{x}_{S,l}) + \sum_{k,l}^{N_B} (x_{B,k} - \bar{x}_{B,k})(x_{B,l} - \bar{x}_{B,l}), \qquad (3)$$

which describes the dispersion of events relative to the means within their own classes, $\bar{x}_S$ and $\bar{x}_B$. Essentially, $C_{S,kl}$ and $C_{B,kl}$ are just covariance matrices of the respective input variables with respect to their signal or background classes.

The Fisher discriminant method minimises the within-class dispersion, where outputs from the same class are confined in a close vicinity while pushing the outputs of different classes as far away as possible from each other. Despite its simplicity, the Fisher discriminant works rather well particularly for Gaussian-distributed input variables with linear correlations; however, it will fail if the input variables have the same sample mean for signal and background, thus making variable transformations an important pre-processing step for this classifier.

An artificial neural network (ANN) analyses and processes information in the same way the human brain works (Reza 2021). It can be considered as a mapping of neurons, performed by calculating the weighted sum of a collection of response functions between a set of input parameters and one or more output parameters (Sadeh et al. 2016). In solving classification problems, the output is in the range of 0 to 1, discriminating between a signal and a background. Sadeh et al. (2016) implemented the TMVA method for ANNs called the multilayer perceptron (MLP, structure shown in **Figure 2**) in ANNz2, where the neurons are organised into at least three layers: input, hidden, and output. The number of layers depend on the complexity of the structure. In this work, we constructed a simple ANN with two hidden layers of 8 and 4 nodes each, each node uses the tanh activation function, and the Broyden-Fletcher-Goldfarb-Shannon (BFGS, Xie et al. 2020) method is employed for backpropagation. An average result based on four runs with different random seeds is taken. Details on the configuration options mentioned can be found at Hoecker et al. (2007).

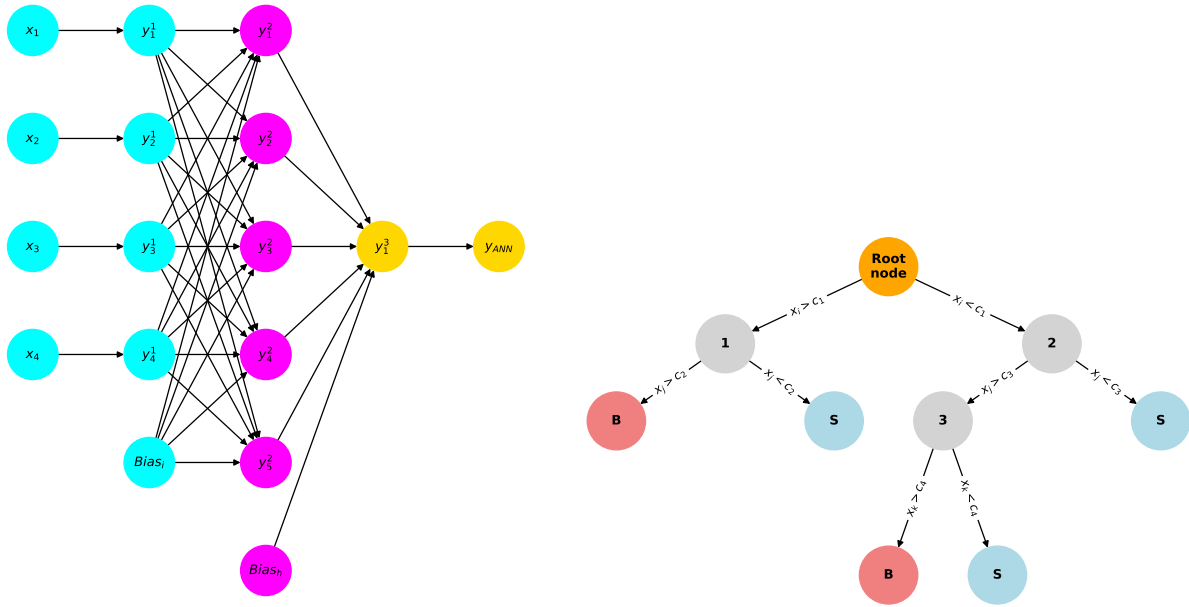

**Figure 2**. The structure of a multi-layer perceptron neural network (left), and a boosted decision tree (right). Figures reproduced from Hoecker et al. (2007).

A boosted decision tree (BDT) is a binary tree where decisions are performed a sample at a time until a stop criterion is fulfilled (Sadeh et al. 2016). Starting from a root node, the sample follows a series of binary splits called internal nodes until a leaf node is reached (Reza 2021). This structure is shown in **Figure 2**. The sample is checked by both the root and internal nodes, then is assigned to the branch corresponding to it. This process is repeated at each new internal node until the sample reaches the leaf node, implying that the sample is assigned to a label. For classification, a leaf is interpreted as either a signal or background. In this work, a BDT is configured with 500 trees in a forest, using bagging with a 1:1 size ratio, and a minimum of 2% of training events required a leaf node. An average result based on four runs with different node splitting is taken.

The *k*-nearest neighbour (*k*NN) algorithm operates on the principle of closeness, that similar events tend to exist close to each other. Classification is performed by looping over all stored events and finding the *k*-nearest neighbours, i.e. events that are closest to the test event.

The parameter $k$ represents the number of nearest neighbours to consider for making predictions. Distance $d_k$ is measured using a metric function called the Euclidean distance,

$$d_k = \sqrt{\sum_{i=1}^{n} |x_i - y_{i,k}|^2}, \tag{4}$$

where $x_i$ and $y_{i,k}$ are the data points from the training and test samples, for $n$ input variables. The $k$ events with the smallest distance are the $k$-nearest neighbours. In this work, we setup a simple $k$NN classifier which uses a Gaussian kernel with a scale fraction set to 0.9. An average result based on four runs with different values of $k$ is taken.

*Pre-Processing Variable Transformation*

Input parameters can be pre-processed for three purposes: to reduce correlations among the parameters, to transform their shapes into more appropriate forms, or to accelerate the response time of a machine learning algorithm (Hoecker et al. 2007). TMVA allows the use of five different variable transformations on input parameters prior to training, they are

1. *Min-max normalisation*, which scales the features to values between 0 to 1. Normalisation reduces the standard deviation for variables in the input space, but does not translate well for features with significant outliers;
2. *Decorrelation*, which produces variables that are not linearly correlated. Decorrelation only works well under two conditions: if the variables were originally linearly correlated, and if they follow the Gaussian distributions, otherwise it will increase the degree of nonlinearity between variables, making the training ineffective. Decorrelation makes the input features contribute independently, reducing overfitting in models;
3. *Principal component analysis* (PCA), which reduces the high dimensionality of complex and large data sets by focusing on the first few principal components (or largest eigenvalues) while ignoring the rest of the variables. PCA works best on data with a large number of input features, it can reduce computation time while retaining important information, allowing the model to work on lower-dimensional and less-correlated data, which prevents overfitting;
4. *Uniformisation*, which uses the cumulative distribution function obtained from the training data to transform the parameters into a uniform distribution;
5. *Gaussianisation*, which transforms the input variables into a Gaussian distribution with zero mean and unity width. Gaussianisation makes the training less affected by outlier input data.

In this work, on top of comparing the training performances between Fisher discriminant, ANN, BDT and $k$NN, we also analyse the effects of normalisation (N), decorrelation (D), PCA (P), uniformisation (U), and Gaussianisation (G) on each of the classifiers, in comparison with no variable transformation (X).

*Performance Metrics*

Performance metrics are measures used to evaluate the overall effectiveness of the discriminants where they provide insights into how well that specific discriminant is performing in classification. In this work, we utilise the confusion matrix and Gaussian kernel density estimation to select the best discriminant by evaluating the overall performance.

A *confusion matrix* provides a visual summary of the model's performance by comparing the actual outcomes with predictions made by the model. The confusion matrix in a binary classification problem is a $2 \times 2$ matrix,

$$\begin{bmatrix} \text{True Positive (TP)} & \text{False Negative (FN)} \\ \text{False Positive (FP)} & \text{True Negative (TN)} \end{bmatrix}. \tag{5}$$

De Diego et al. (2022) provided a list of performance metrics that can be obtained from a confusion matrix. The scores of all the metrics range from 0.0 to 1.0, where 1.0 is the highest possible score. The four metrics used in this research are accuracy, precision, recall and F1 score. *Accuracy* is used to evaluate the overall proportion of both true positives and true negatives for its performance in classification,

$$\text{Accuracy} = \frac{\text{TP} + \text{TN}}{\text{TP} + \text{TN} + \text{FP} + \text{FN}} = \frac{\text{T}}{\text{P} + \text{N}}, \tag{6}$$

where $\text{T} = \text{TP} + \text{TN}$ (total true), $\text{P} = \text{TP} + \text{FP}$ (total positives), and $\text{N} = \text{TN} + \text{FN}$ (total negatives). The *precision* measures the accuracy of how the model classifies the positive predictions of both true and false positives,

$$\text{Precision} = \frac{\text{TP}}{\text{TP} + \text{FP}}. \tag{7}$$

*Recall* guarantees the model's sensitivity in identifying positive occurrences to determine whether the model can classify the signal better than the background or inversely,

$$\text{Recall} = \frac{\text{TP}}{\text{TP} + \text{FN}}. \tag{8}$$

Finally, a high F1 score guarantees that the model must function well in terms of both recall and precision. The proportion and balance between recall and precision can be seen in the F1 score,

$$\text{F1 score} = 2 \times \frac{\text{precision} \times \text{recall}}{\text{precision} + \text{recall}}. \tag{9}$$

Gaussian Kernel Density Estimation (KDE) is used to visualise the performance of each discriminant. Gaussian KDE plots each data point to visualise the distribution of the data according to their predicted classes and produces a smooth curve of the probability density function for the total data. From this, we may learn about the characteristics of each discriminant and how biased or effective they are at classifying both positive and negative instances by displaying the spreads, densities, and overlaps of curves. The region where the discriminant is uncertain about their ability to distinguish between the classes is represented by the overlap of curves. A decision threshold of 0.5 was chosen for this study, which is a vertical line that establishes the classification cutoff. In other words, this threshold is the transition line of one class to another (Węglarczyk 2018).

## Results and Discussion

### *Variable Transforms and Their Effects on the Fisher Discriminant*

By comparing the performance metrics of the 24 runs performed (6 variable transformations × 4 machine learning algorithms, see **Figure 3**), we found that the transformations have different effects on the performance of the tested algorithms. We summarise and tabulate the results in **Table 1**. Focusing on the Fisher discriminant, uniformisation and Gaussianisation improved the accuracy, having a difference of 0.8% and 0.76% respectively compared to the default run (i.e. no variable transformation). The same effects can be seen with the precision (1.48% and 2.9% improvement) and F1 score (0.76% and 0.59% improvement). However, Gaussianisation reduced the recall score by 1.55%.

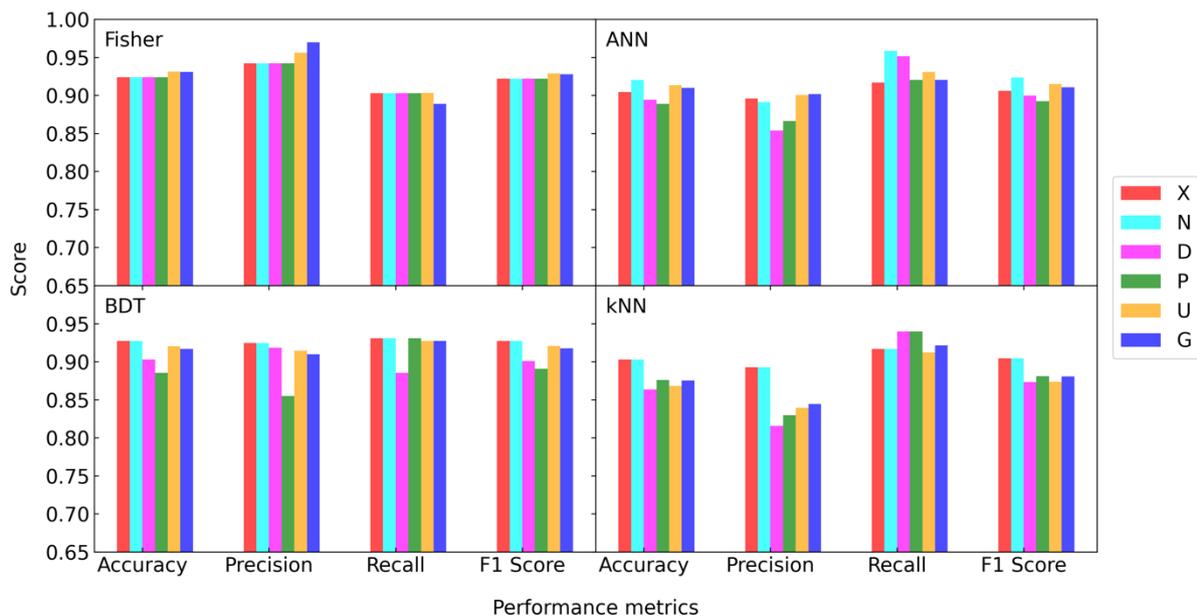

**Figure 3**. A comparison of the F1 score of five pre-processing variable transformations; normalisation, decorrelation, principal component analysis (PCA), uniformisation, and Gaussianisation, across the performance metrics for the four ML algorithms Fisher discriminants (top left), artificial neural networks (ANN, top right), boosted decision trees (BDT, bottom left) and *k*-nearest neighbours (*k*NN, bottom right).

As for ANN, uniformisation and Gaussianisation improved its performance with an improvement of 0.96% and 0.57% for accuracy, 0.53% and 0.66% for precision, 1.5% and 0.38% for recall, and 1.01% and 0.54% for the F1 score, respectively. Normalisation also significantly improved the accuracy, recall, and F1 score with an improvement of 1.71%, 4.44%, and 1.91% respectively. The precision, however, was reduced with a 0.54% difference.

For the remaining two transformations, decorrelation and PCA, only the recall scores were improved, having differences of 3.72% and 0.38% respectively compared to the default.

**Table 1**. Performance metrics and their respective uncertainties ($\sigma$) of six pre-processing transformations: no transformation (X), normalisation (N), decorrelation (D), principal component analysis (P), uniformisation (U), and Gaussianisation (G), for the four ML algorithms Fisher discriminant, artificial neutral network (ANN), boosted decision tree (BDT) and k-nearest neighbour (*k*NN). Entries highlighted in green indicates the highest value within each variable transform, while those in blue indicate the highest value of each metric.

| Algorithm | Accuracy | $\sigma$ | Precision | $\sigma$ | Recall | $\sigma$ | F1 Score | $\sigma$ |
|---|---|---|---|---|---|---|---|---|
| **X** | | | | | | | | |
| Fisher | 0.9236 | | 0.9420 | | 0.9028 | | 0.9220 | |
| ANN | 0.9045 | 0.0077 | 0.8958 | 0.0145 | 0.9167 | 0.0098 | 0.9058 | 0.0071 |
| BDT | 0.9271 | 0.0045 | 0.9244 | 0.0082 | 0.9306 | 0.0000 | 0.9274 | 0.0041 |
| *k*NN | 0.9027 | 0.0106 | 0.8927 | 0.0188 | 0.9166 | 0.0000 | 0.9043 | 0.0095 |
| **N** | | | | | | | | |
| Fisher | 0.9236 | | 0.9420 | | 0.9028 | | 0.9220 | |
| ANN | 0.9201 | 0.0083 | 0.8910 | 0.0141 | 0.9583 | 0.0000 | 0.9233 | 0.0075 |
| BDT | 0.9271 | 0.0045 | 0.9244 | 0.0082 | 0.9306 | 0.0000 | 0.9274 | 0.0041 |
| *k*NN | 0.9027 | 0.0106 | 0.8927 | 0.0188 | 0.9166 | 0.0000 | 0.9043 | 0.0095 |
| **D** | | | | | | | | |
| Fisher | 0.9236 | | 0.9420 | | 0.9028 | | 0.9220 | |
| ANN | 0.8941 | 0.0017 | 0.8537 | 0.0046 | 0.9514 | 0.0040 | 0.8998 | 0.0013 |
| BDT | 0.9028 | 0.0063 | 0.9184 | 0.0130 | 0.8854 | 0.0183 | 0.9009 | 0.0071 |
| *k*NN | 0.8635 | 0.0101 | 0.8158 | 0.0141 | 0.9398 | 0.0046 | 0.8732 | 0.0083 |
| **P** | | | | | | | | |
| Fisher | 0.9236 | | 0.9420 | | 0.9028 | | 0.9220 | |
| ANN | 0.8889 | 0.0085 | 0.8664 | 0.0119 | 0.9201 | 0.0066 | 0.8924 | 0.0078 |
| BDT | 0.8854 | 0.0134 | 0.8550 | 0.0204 | 0.9306 | 0.0000 | 0.8908 | 0.0113 |
| *k*NN | 0.8760 | 0.0096 | 0.8296 | 0.0195 | 0.9398 | 0.0046 | 0.8810 | 0.0103 |
| **U** | | | | | | | | |
| Fisher | 0.9310 | | 0.9560 | | 0.9030 | | 0.9290 | |
| ANN | 0.9132 | 0.0083 | 0.9006 | 0.0175 | 0.9306 | 0.0057 | 0.9149 | 0.0071 |
| BDT | 0.9201 | 0.0035 | 0.9146 | 0.0076 | 0.9271 | 0.0035 | 0.9207 | 0.0031 |
| *k*NN | 0.8681 | 0.0120 | 0.8392 | 0.0185 | 0.9120 | 0.0046 | 0.8738 | 0.0101 |
| **G** | | | | | | | | |
| Fisher | 0.9306 | | 0.9697 | | 0.8888 | | 0.9275 | |
| ANN | 0.9097 | 0.0049 | 0.9018 | 0.0086 | 0.9201 | 0.0104 | 0.9106 | 0.0050 |
| BDT | 0.9167 | 0.0049 | 0.9097 | 0.0173 | 0.9271 | 0.0104 | 0.9177 | 0.0037 |
| *k*NN | 0.8750 | 0.0120 | 0.8445 | 0.0210 | 0.9213 | 0.0046 | 0.8808 | 0.0097 |

Despite all the positive effects of transformations seen in Fisher and ANN, the same cannot be said for BDT. All but the normalisation resulted in poorer performance, having the performance metric values decrease with the transformations applied. Normalisation did not improve the scores, but only kept it static. The negative effects of the transformations on BDT can be seen in *k*NN as well. Decorrelation and PCA improved the recall for *k*NN compared to the default, with differences of 2.5% for both, and 0.5% for Gaussianisation.

Therefore, we conclude that the impact of pre-processing variable transformations on the performance of the tested algorithms varies significantly with respect to each machine learning method. The Fisher discriminant and ANN benefited the most from uniformisation and Gaussianisation, showing improvements in accuracy, precision, and F1 scores. In contrast,

BDT and *k*NN showed either no changes or degradations in performance metric with most transformations. These findings emphasise the importance of selecting suitable pre-processing techniques based on the specific algorithm, showing that proper pre-processing of data is a must for Fisher discriminants as well as ANNs, while this process can be ignored when using BDTs and *k*NNs.

*Comparison between the Fisher Discriminant with ANN, BDT and kNN*

Taking the highest scores for each performance metric, we find that the Fisher discriminant with uniformisation transformation performs the best, achieving the highest accuracy and F1 score. The performance gap between Fisher and the other models are 1.93%, 0.42%, and 3.08% for accuracy, and 0.62%, 0.17%, and 2.69% for the F1 score, when compared to ANN, BDT, and *k*NN, respectively. In terms of precision, Fisher discriminant with Gaussianisation excels, surpassing ANN, BDT, and *k*NN by 7.26%, 4.79%, and 8.27%, respectively. However, for recall, ANN with normalisation takes the lead, outperforming the Fisher discriminant by 0.62%, BDT by 0.17%, and *k*NN by 2.69%.

Going into the specifics, we show the Gaussian KDE plots of each pre-processing variable transformation in **Figure 4**, where we see that Fisher's classification of no-bulge galaxy consistently exhibits high density peaks as compared to ANN, BDT and *k*NN across all the transformations, indicating that the Fisher discriminant is highly accurate in classifying no-bulge galaxies. However, for rounded galaxies, the discriminant consistently shows the lowest density across all transformations, this situation is reflected by its recall values which are generally lower than all other algorithms due to the high count of false negatives (i.e. rounded galaxies which are misclassified as no-bulge galaxies). This means that if the research goal is to specifically filter and discard rounded galaxies from a data set, ANN is the preferred algorithm as compared to Fisher. At the same time, ANN and BDT exhibit higher densities for the classification of rounded galaxies compared to *k*NN and the Fisher discriminant. While *k*NN performs similarly to the Fisher discriminant in classifying rounded galaxies, there are more overlaps between the two galaxy types, particularly in transformations N, U, and G, implying that *k*NN struggles to distinguish between the classes. Although ANN and BDT are less accurate in classifying no-bulge galaxies, they offer more flexibility by exhibiting nearly equal density for both galaxy types, as seen in X and transformations U and G.

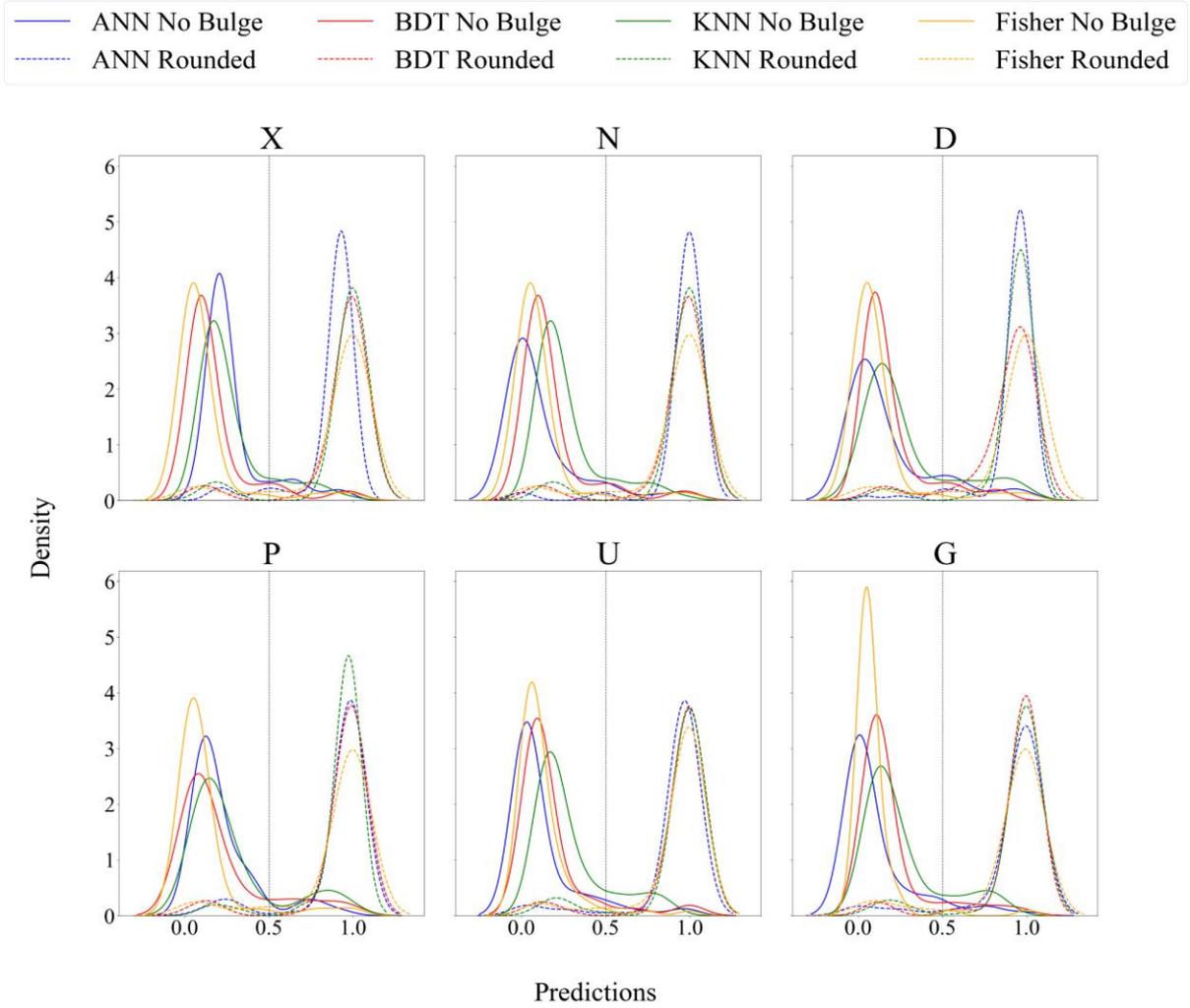

**Figure 4**. Gaussian KDE graph for no transformation (X, top left), normalisation (N, top middle), decorrelation (D, top right), principal component analysis (P, bottom left), uniformisation (U, bottom middle) and Gaussianisation (G, bottom right).

Hence, it can be concluded that the Fisher discriminant has shown a very good performance, particularly in handling simple classification tasks. This advantage may stem from the complexity of the other models, where every adjustment to the algorithm can significantly impact their outcomes. In contrast, Fisher's simpler structure allows it to maintain a more consistent and reliable performance. As an example, **Figure 5** shows how the Gaussian KDE curves vary significantly with different seed numbers for an ANN training: an ANN with random seed value 44 exhibits half the density for rounded galaxies compared to an ANN with random seed value 444. Moreover, the ANN with random seed value 44 outputs predictions closer to theoretical values, while the ANN with random seed 444 outputs predictions closer to the boundary line of 0.5, increasing the overlap between classes. This shows that ANN is highly sensitive to seed numbers (consistent with the findings of Mahmud Pathi et al. 2025), which can lead to inconsistent performance. Selecting an appropriate seed for ANN requires substantial time and effort, as we need to test and compare many different seeds to produce the best classification performance; whereas the Fisher discriminant is shown to produce robust results within a single run in a short amount of time. As an example, our Fisher discriminant algorithm took 40 seconds to produce our results, whereas ANN needs 5 minutes.

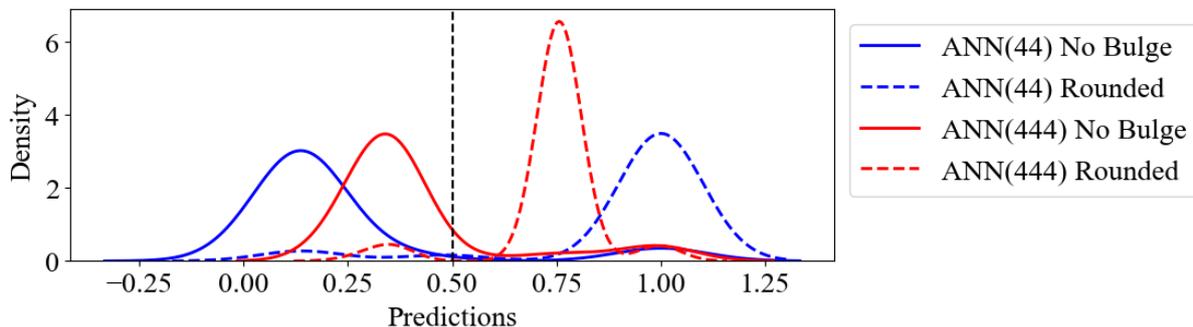

**Figure 5**. Gaussian KDE graph for an ANN with random seed values of 44 and 444 in a no-transformation run (X), highlighting the dependence of an ANN on its hyperparameters.

**Conclusion**

In this research, we explored the viability of the Fisher discriminant in performing simple classification tasks by applying pre-processing variable transformations and comparing its performance with three well-known classifiers: ANN, BDT, and *k*NN. The impact of variable transformations varied depending on the algorithm and transformation, sometimes improving, having no effect, or even worsening the performance. In terms of accuracy, the Fisher discriminant with uniformisation outperformed all other algorithms, achieving the highest accuracy (0.9310, an improvement of 0.4% as compared to the next best algorithm). For precision, Fisher with Gaussianisation achieved the highest score (0.9697, an improvement of 4.9%) while ANN with normalisation had the best recall (0.9583). Overall, Fisher with uniformisation has the highest F1 score (0.9290, an improvement of 0.2%), demonstrating excellent performance, especially for tasks with small sample sizes. We believe that ANN, BDT and *k*NN are sensitive to the complex hyperparameter dependency within each algorithm; unlike the Fisher discriminant that is more straightforward, as it doesn't require parameters tuning and is relatively simple to implement, thus making the overall training process less time consuming while still delivering competitive results.

Further analysis on the viability of the Fisher discriminant could explore a wider range of input parameters, as SDSS provides many other variables that have yet to be tested. While our results show significant promise, it is too early to consider the Fisher discriminant as the definitive solution to all classification problems, since we believe that the analysis might be sample dependent: our results are only valid for a low redshift galaxy sample such that of the SDSS. Although the absolute improvement of accuracy and F1-score compared to other algorithms is low (below 1%), we believe that the Fisher discriminant has great potential due to its simple structure and short running time as compared to all other algorithms tested in this work. We also believe that Fisher's recall can be improved if a more thorough feature selection methodology was conducted, especially exploring the galaxy properties which could highlight the presence of a galaxy's bulge.

In future work, feature selection methods could be incorporated, testing can be extended to other branches of the GZ2 decision tree or other data samples to improve the current results. The research in classifying celestial sources remains a very relevant and essential field to pursue in the near future, we hope that this work will ultimately contribute to the quest of better understanding the universe we live in.

## Acknowledgements

JYHS acknowledges financial support via the Fundamental Research Grant Scheme (FRGS) by the Malaysian Ministry of Higher Education with code FRGS/1/2023/STG07/USM/02/14.